\documentclass[12pt]{iopart}

\usepackage{graphics}
\usepackage{amsmath}
\usepackage[dvips]{graphicx}
\usepackage{amsfonts,amsbsy}
\usepackage{amssymb}
\usepackage{bbm}
\newcommand{\beq}{\begin{equation}}
\newcommand{\eeq}{\end{equation}}
\newcommand{\beqa}{\begin{eqnarray}}
\newcommand{\eeqa}{\end{eqnarray}}

\def\k{{\boldsymbol k}}
\def\t{{\boldsymbol t}}
\def\q{{\boldsymbol q}}

\def\0{{\boldsymbol 0}}

\def\K{{\boldsymbol K}}

\def\co{{\mathcal C}_0}
\def\ci{{\mathcal C}_1}
\def\cii{{\mathcal C}_2}

\begin{document}
\title{Suppression of high-$p_T$ particle production in AA collisions: the role of in-medium color-flow}
\author{A. Beraudo$^{1,2}$, J.G. Milhano$^{2,3}$ and U.A. Wiedemann$^2$}
\address{$^1$Centro Studi e Ricerche ``Enrico Fermi'', Piazza del Viminale 1, Roma, Italy}
\address{$^2$Physics Department, Theory Unit, CERN, CH-1211 Gen\`eve 23, Switzerland}
\address{$^3$CENTRA, Instituto Superior T\'ecnico, Universidade T\'ecnica de Lisboa,\\ Av. Rovisco Pais 1, P-1049-001 Lisboa, Portugal} 
\ead{Andrea.Beraudo@cern.ch, guilherme.milhano@ist.utl.pt, Urs.Wiedemann@cern.ch}
\begin{abstract}
The suppression of high$-p_T$ single-hadron spectra in heavy-ion collisions is usually interpreted as due to 
parton energy-loss of high-momentum quarks and gluons propagating in the plasma. 
Here, we discuss to what extent this partonic picture must be 
complemented by a picture of medium-modified hadronization. In particular, we show how color-exchange with the medium modifies the properties of color singlet-clusters arising from the parton branchings, producing a softening of the hadron spectra.    
\end{abstract}
Suppression of high-momentum particle spectra in nucleus-nucleus collisions, a.k.a. 
\emph{jet-quenching}, discovered at RHIC and now explored at the LHC, signaled that in 
heavy-ion experiments a very dense and opaque medium is produced.
The observed suppression is usually attributed to energy-loss at the partonic level: high-momentum partons produced in hard processes lose energy due to collisions and/or gluon-radiation in the plasma, thus giving rise, at hadronization, to a softer hadron spectrum.

Such a picture relies on the factorization assumption
\beq
d\sigma_{\rm med}^{AA\to h+X}=\sum_f d\sigma_{\rm vac}^{AA\to f+X}\otimes{{\langle P(\Delta E)\rangle_{AA}}}\otimes{{D_{\rm vac}^{f\to h}(z)}}\, ,
\label{eq:fact}
\eeq
that expresses the hadron production in AA collisions through the convolution of a hard partonic cross section with a medium-induced energy-loss probability distribution and a \emph{vacuum} fragmentation function. High-$p_T$ partons are in general expected to hadronize \emph{outside the medium}, essentially due to Lorentz time-dilatation. A parton of virtuality $Q$ hadronizes in its rest frame within a time $\Delta t^{\rm rest}\!\sim\!1/Q$. In the lab-frame such a time is dilated to $\Delta t^{\rm lab}\!\sim\!E/Q^2$. This motivates the standard assumption that  for large enough energy $E$, hadrons should be produced in the vacuum and Eq.~(\ref{eq:fact}) should apply. Here we point out
that color-exchange between the hard parton and the plasma introduces effects which cannot be captured by the this simple factorized ansatz.

\begin{figure}[!tp]
\begin{center}
\includegraphics[clip,width=0.48\textwidth]{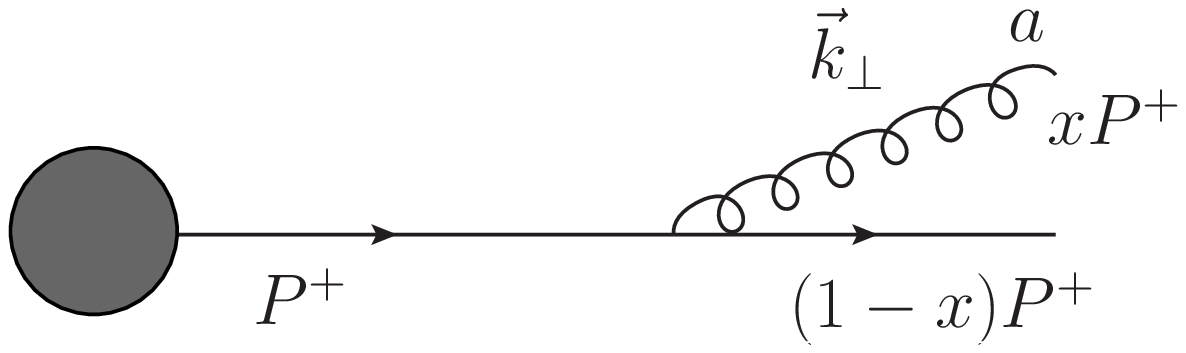}
\includegraphics[clip,width=0.48\textwidth]{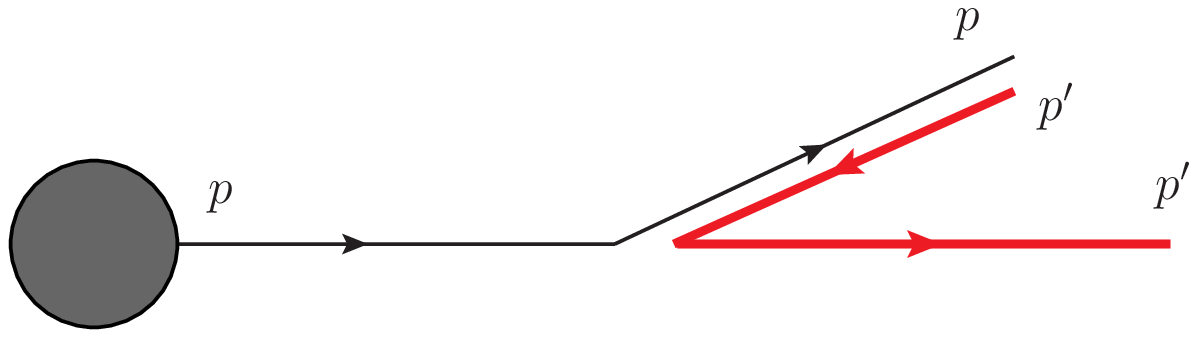}
\caption{Vacuum radiation of a hard parton with the associated color-flow. In red the cluster $\co$, giving rise to the leading hadron.} 
\label{fig:vacuum}
\end{center}
\end{figure}
In elementary collisions hadronization occurs at the final stage of a parton shower evolution, through the decay of color-singlet clusters. Let us start then by considering the branching of an off-shell parton (R) produced in a hard event which occurred in the vacuum: this will represent our benchmark for the study of medium effects. The spectrum of radiated gluons has the well-known expression (in light-cone coordinates):
\beq
d\sigma^{\rm vac}=d\sigma^{\rm hard}\frac{\alpha_s}{\pi^2}C_R{\frac{dk^+}{k^+}}{\frac{d\k}{\k^2}},\quad\quad k=\left[k^+,\frac{\k^2}{2k^+},\k\right].
\eeq
Considering color-flow (in the large-$N_c$ limit) one can split the gluon into a $q\overline{q}$ pair and identify in Fig.~\ref{fig:vacuum} the cluster $\co$ with momentum $p_{\co}\!=\!p_f\!+\!k/2$, so that
\beq
p_{\co}^+=(1-x/2)\,p^+,\qquad M_{\co}^2=\k^2/2x(1-x),\label{eq:vacuum}
\eeq
whose decay will give rise to hadrons.

We now consider how the presence of a medium affects the branching pattern of a high-energy parton, both at the level of the spectrum of radiated gluons and of the properties of color-singlet clusters. The analysis will be performed at the $N\!=\!1$ order of the opacity expansion, i.e. considering a single scattering with the medium. Notice that, for the purpose of facing single-particle spectra, we can focus on the medium effects on \emph{leading hadrons}. For a steeply-falling parton spectrum $dN_f/dp_T\!\sim\!p_T^{-n}$ one has, in fact, at the hadron level $dN_h/p_T\!\sim\!p_T^{-n}\int_0^1dz\, z^{n-1}D_f^h(z)$, so that only particles carrying a significant fraction of the parton momentum will contribute to the spectrum. 

\begin{figure}[!tp]
\begin{center}
\includegraphics[clip,width=0.8\textwidth]{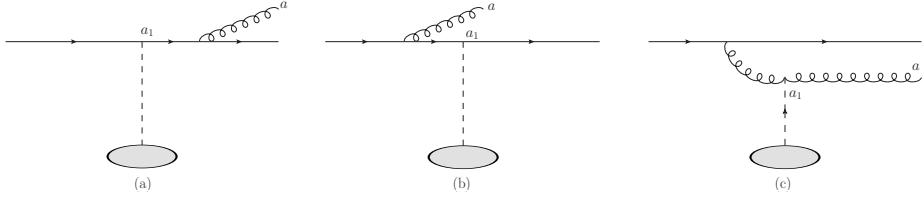}
\caption{The three diagrams contributing to the medium-induced radiation spectrum of a hard on-shell parton at the lowest order in opacity.} 
\label{fig:N1}
\end{center}
\end{figure}
\begin{figure}[!tp]
\begin{center}
\includegraphics[clip,width=0.49\textwidth]{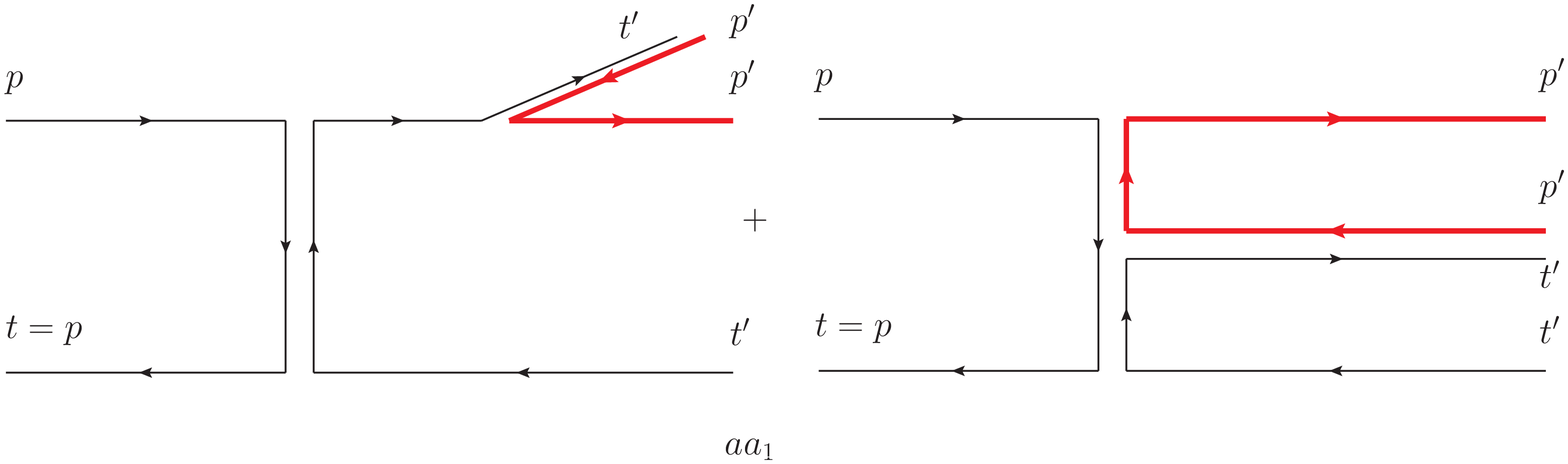}\includegraphics[clip,width=0.49\textwidth]{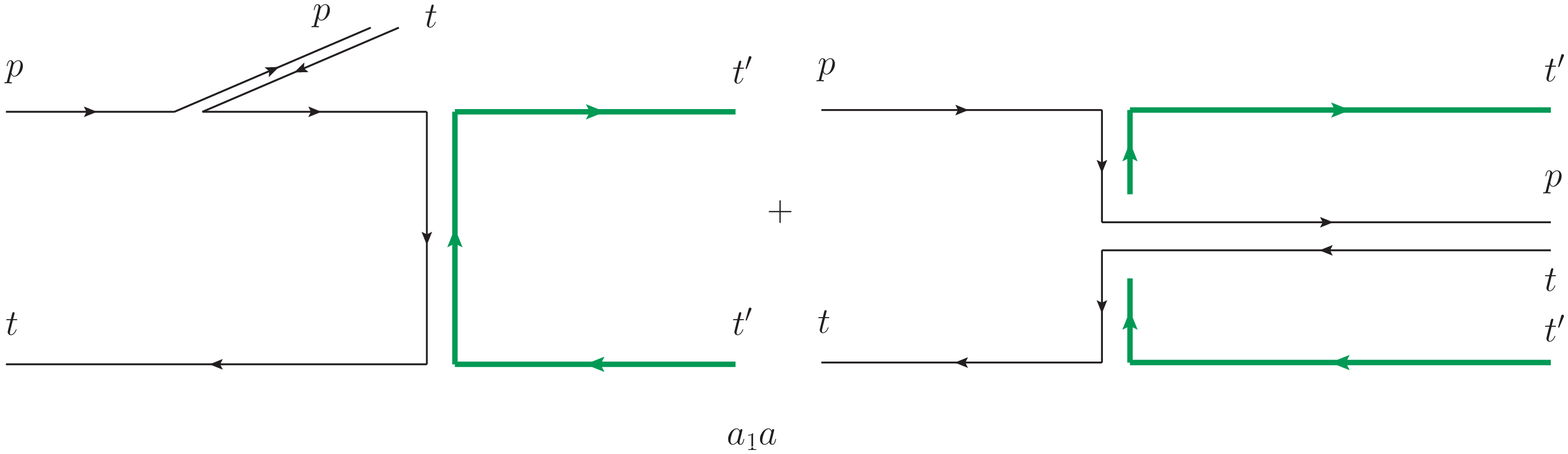}
\caption{The two different color-flows involved in the radiation spectrum at the $N\!=\!1$ order in opacity. The three-gluon vertex contributes to both color channels.} 
\label{fig:N1_color}
\end{center}
\end{figure}

We first consider the case of an on-shell parton produced far in the past. The relevant diagrams are shown in Fig.~\ref{fig:N1}. As displayed in Fig.~\ref{fig:N1_color}, the spectrum can be decomposed into two independent color channels, $aa_1$ and $a_1a$, which are found to give an identical contribution and summed together provide the usual Gunion-Bertsch spectrum:
\beq
\left.k^+\frac{dN_g}{d\k dk^+}\right|_{aa_1}=\left.k^+\frac{dN_g}{d\k dk^+}\right|_{a_1a}=\frac{N_c}{2}\frac{\alpha_s}{\pi^2}\left\langle\left[\K_0-\K_1\right]^2\right\rangle.
\eeq
In the above $\K_0\!\equiv\!\k/\k^2$, $\K_1\!\equiv\!(\k\!-\!\q)/(\k\!-\!\q)^2$ and the momentum $\q$ exchanged with the medium  is weighted by the corresponding differential elastic cross-section.
Though the radiation spectrum in the two channels is the same, the properties of the clusters giving rise to the leading hadron are very different.
In the $aa_1$ case (left panel of Fig.~\ref{fig:N1_color}) the cluster $\ci$ is characterized by the momentum $p_{\ci}^+\!=\!(1\!-\!x/2)p^+$, and the invariant mass $M_{\ci}^2\!\underset{x\ll 1}{\sim}\!\k^2/2x$,
hence sharing the same kinematics of vacuum radiation described in Eq.~(\ref{eq:vacuum}).
On the other hand in the $a_1a$ channel the cluster $\cii$ (right panel of Fig.~\ref{fig:N1_color}) arises from the combination of the leading fragment of the projectile with a parton of the medium $t\!\equiv\![t^+,\t^2/2t^+,\t]$, with typical momentum components of order of the temperature $T$. One has then $p_{\cii}^{+}\!\approx\!(1\!-\!x)p_+$ and $M_{\cii}^2\!\equiv\!\left(p_f\!+\!t\right)^2\!\approx\!2(1-x) p_+{\t^2}/{2t_+}$.
Being $M_{\cii}^2\!\sim\! ET$, for large enough parton energy one has $M_{\cii}\!\gg\!M_{\ci}$. Hence the cluster $\cii$ will give rise to a softer gluon spectrum: first because of its softer ``$+$''-momentum; secondly because of its larger invariant mass, which will lead in the decay to a larger hadron multiplicity. The above argument is completely general. Here for illustration we focus on one particular model: the cluster-hadronization employed by HERWIG~\cite{herwig}. Light clusters ($M_{\mathcal C}\!<\!M_c\!=\!4$ GeV) suffer a two-body decay into hadrons; for values of the momenta of phenomenological interest this is the case for $\ci$. Heavy clusters ($M_{\mathcal C}\!>\!M_c$; in elementary collisions they are very rare!) first suffer a fission into two sub-clusters, which are then made decay into hadrons; this is typically the case for $\cii$. The leading sub-cluster $X$ will carry momentum ($Q_0\!=\!0.6$ GeV is a parameter of HERWIG)
\beq
p_X=\left(1-Q_0/M_{\cii}\right)p_f+Q_0/M_{\cii}t\approx\left(1-Q_0/M_{\cii}\right)p_f.
\eeq
Hence hadron momenta in the $a_1a$ channel will pay a penalty factor:
\beq
p_X^+/p_{\ci}^+\approx(1-x)\left(1-Q_0/M_{\cii}\right)/(1-x/2)
\eeq
which, on top of a steeply falling parton spectrum, can give a sizable effect on the nuclear modification factor $R_{AA}$.    

We now consider the more realistic situation in which the hard parton is produced in the medium at a finite $t_0$. This introduces new conceptual issues: the first one is to distinguish the medium-induced radiation ($d\sigma^{\rm ind}$) from the one that would occur also in the vacuum ($d\sigma^{\rm vac}$); secondly the finite separation between the production and the scattering vertexes sets a time-scale to compare with the \emph{formation-time} of the radiated gluon. One has then, considering also the color-flow:
\beq
d\sigma^{\rm rad}\equiv{d\sigma^{\rm vac}+d\sigma^{\rm ind}}={d\sigma_a^{\rm vac}+d\sigma_a^{\rm ind}}+{d\sigma_{aa_1}^{\rm ind}}+{d\sigma_{a_1a}^{\rm ind}},
\eeq
where $d\sigma_a^{\rm ind}$ is a correction of the vacuum-radiation spectrum to ensure unitarity (the radiation must be accompanied either by zero or one elastic interaction). Two time-scales are found to be relevant for the study of color-flow: $1/\omega_1^-\!\equiv\! 2k^+/{(\k\!-\!\q)^2}$ and $1/\omega_0^-\!\equiv\! 2k^+/{\k^2}$ depending on the initial/final gluon transverse momentum. They have to be compared with the medium `length' $L^+$.  
For details we refer the reader to~\cite{BMW}. Here we simply wish to illustrate the result of the two limiting cases.
The first one is the \emph{totally incoherent} regime, when $1/\omega_{1}^-,1/\omega_0^-\!\ll\!L^+$. One has:
{\setlength\arraycolsep{1pt}
\beqa
\left.k^+\frac{d\sigma^{\rm rad}}{dk^+ d\k}\right|_{aa_1}^{\rm incoh.}
&=&d\sigma^{\rm hard}\frac{\alpha_s}{2\pi^2}\frac{L^+}{\lambda_{\rm el}^+(R)}\,C_A\left\langle\left(\K_0\!-\!\K_1\right)^2+\K_1^2\right\rangle\nonumber\\
\left.k^+\frac{d\sigma^{\rm rad}}{dk^+ d\k}\right|_{a_1a}^{\rm incoh.}&=&d\sigma^{\rm hard}\frac{\alpha_s}{2\pi^2}\!\frac{L^+}{\lambda_{\rm el}^+(R)}\,C_A\left\langle(\K_0\!-\!\K_1)^2+\K_0^2+\K_1^2\right\rangle\nonumber\\
\left.k^+\frac{d\sigma^{\rm rad}}{dk^+ d\k}\right|_{a}^{\rm incoh.}&=&k^+\frac{d\sigma^{\rm vac}}{dk^+ d\k}+d\sigma^{\rm hard}\frac{\alpha_s}{2\pi^2}\frac{L^+}{\lambda_{\rm el}^+(R)}\,C_A(-3\K_0^2)
\eeqa}
For the color-inclusive \emph{induced} spectrum one gets then $d\sigma^{\rm ind}\!\sim\!\langle(\K_0-\K_1)^2+\K_1^2-\K_0^2\rangle$; notice that more then 50\% of the induced radiation spectrum comes from the $a_1a$ channel, associated to a $\cii$-like cluster and hence to a softer hadron spectrum.
The opposite case it the \emph{totally coherent} regime, occurring when $1/\omega_{1}^-,1/\omega_0^-\!\gg\!L^+$. One has in this case
{\setlength\arraycolsep{1pt}
\beqa
\left.k^+\frac{d\sigma^{\rm rad}}{dk^+ d\k}\right|_{aa_1}^{\rm coher.}&=&{d\sigma^{\rm hard}\frac{L^+}{\lambda_{\rm el}^+(R)}C_A\frac{\alpha_s}{2\pi^2}\K_0^2},\qquad\left.k^+\frac{d\sigma^{\rm rad}}{dk^+ d\k}\right|_{a_1a}^{\rm coher.}=0,\nonumber\\
\left.k^+\frac{d\sigma^{\rm rad}}{dk^+ d\k}\right|_{a}^{\rm coher.}&=&k^+\frac{d\sigma^{\rm vac}}{dk^+ d\k}{-d\sigma^{\rm hard} \frac{L^+}{\lambda_{\rm el}^+(R)}C_A\frac{\alpha_s}{2\pi^2}\K_0^2},
\eeqa}
so that for the inclusive result $d\sigma^{\rm ind}\!=\!0$: one is left simply with the vacuum-radiation. The only effect of the medium is to produce a color-rotation of the projectile, so that part of the radiation occurs in the $aa_1$ channel: however in the latter hadronization involves a $\ci$-like cluster and occurs just like in the vacuum. The radiation cross-section corresponding to a non-trivial color-flow vanishes exactly: $d\sigma_{a_1a}^{\rm ind}\!=\!0$.

In summary, we have outlined how color-exchange with the medium can affect the branching of high-energy partons propagating in the QGP, giving rise to ``pre-confined'' objects (color-singlet clusters) having different properties with respect to what occurring in elementary collisions. This can have a sizable effect on the single-particle 
spectra~\cite{BMW}, contributing to a softening of the hadron momenta. As an illustration we have considered the medium-induced gluon-radiation within an opacity expansion.

\vspace*{2mm}
\noindent\textbf{Acknowledgments} JGM acknowledges the support of Funda\c c\~ao para a Ci\^encia e a Tecnologia (Portugal) under project CERN/FP/116379/2010.

\end{document}